\documentclass{PoS}

\title{Nonperturbative infrared fixed point in sextet QCD }

\ShortTitle{Nonperturbative infrared fixed point in sextet QCD }

\author{\speaker{Benjamin Svetitsky} and Yigal Shamir%
\\
        Raymond and Beverly Sackler School of Physics and Astronomy, Tel Aviv University, 69978 Tel Aviv, Israel\\
        E-mail: \email{bqs@julian.tau.ac.il}, \email{shamir@post.tau.ac.il}}

\author{Thomas DeGrand\\
       Department of Physics, University of Colorado, Boulder, CO 80309, USA\\
                E-mail: \email{degrand@pizero.colorado.edu}}

\abstract{The SU(3) gauge theory with fermions in the sextet representation is one of several theories of interest
for technicolor models. We have carried out a Schrodinger functional (SF) calculation for the lattice
theory with two flavors of Wilson fermions. We find that the discrete beta function changes sign
when the SF renormalized coupling is in the neighborhood of $g^2 = 2.0$, showing a breakdown of the
perturbative picture even though the coupling is weak. The most straightforward interpretation is an
infrared-stable fixed point.}

\FullConference{The XXVI International Symposium on Lattice Field Theory \\
                 July 14 - 19, 2008\\
                 Williamsburg, Virginia, USA}

\def\eval#1{\left\langle#1\right\rangle}
\def\co{{\cal O}}
\def\tr{{\rm tr}\,}
\begin{document}

\section{Motivation---Beyond the Standard Model}

Most interesting theories that go beyond the Standard Model require nonperturbative information in order to demonstrate their relevance to low-energy physics \cite{Nelson}.
Among these are strong-coupling realizations of the Weinberg--Salam theory; technicolor, in which the Higgs multiplet arises as bound states of a higher-energy theory; and theories based on extra dimensions, in which Kaluza--Klein phenomena may depend on strong-coupling physics.
Moreover, various theories of unification start with large gauge groups that reduce to low-energy theories via tumbling and vacuum alignment, phenomena that are inherently nonperturbative.  Supersymmetry can only apply to the low-energy world if it is broken by some nonperturbative mechanism.

As a first step in what we hope will develop into a broad attack on gauge theories beyond QCD, we have chosen to study the SU(3) gauge theory with two flavors of Wilson fermions in the sextet representation \cite{SSD}.
The two-loop beta function of this theory crosses zero \cite{Caswell,BZ} at $g^2\simeq10.4$, which is a strong coupling; a ladder calculation indicates that the quarks condense and chiral symmetry is spontaneously broken before this coupling is reached.
If indeed chiral symmetry is broken in this theory, it becomes a candidate for a theory of walking technicolor \cite{Belyaev}, assuming that a lattice calculation can confirm the slow evolution of the coupling.
The survival of the fixed point, on the other hand, would put the theory in the conformal window and disqualify it for technicolor.

A related issue is the possibility of scale separation, where the confinement scale of the theory is at a lower energy than the scale of the chiral condensate \cite{Kogut}.
This is what initially attracted us to studying quarks in higher representations than the fundamental.

\section{Perturbative renormalization group}

\begin{figure}[hbt]
\begin{center}
\includegraphics[width=.4\textwidth]{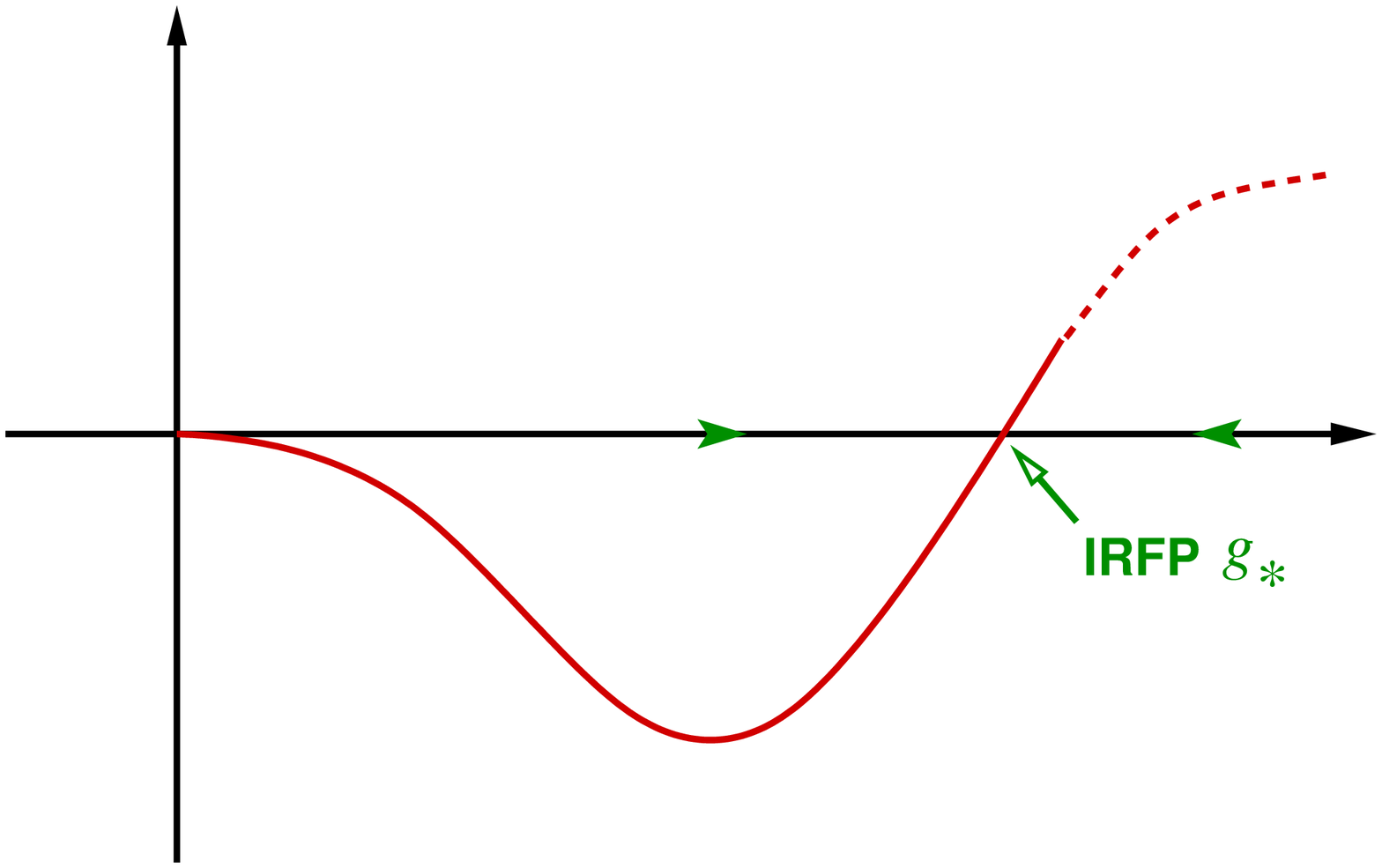}$\quad$
\includegraphics[width=.5\textwidth]{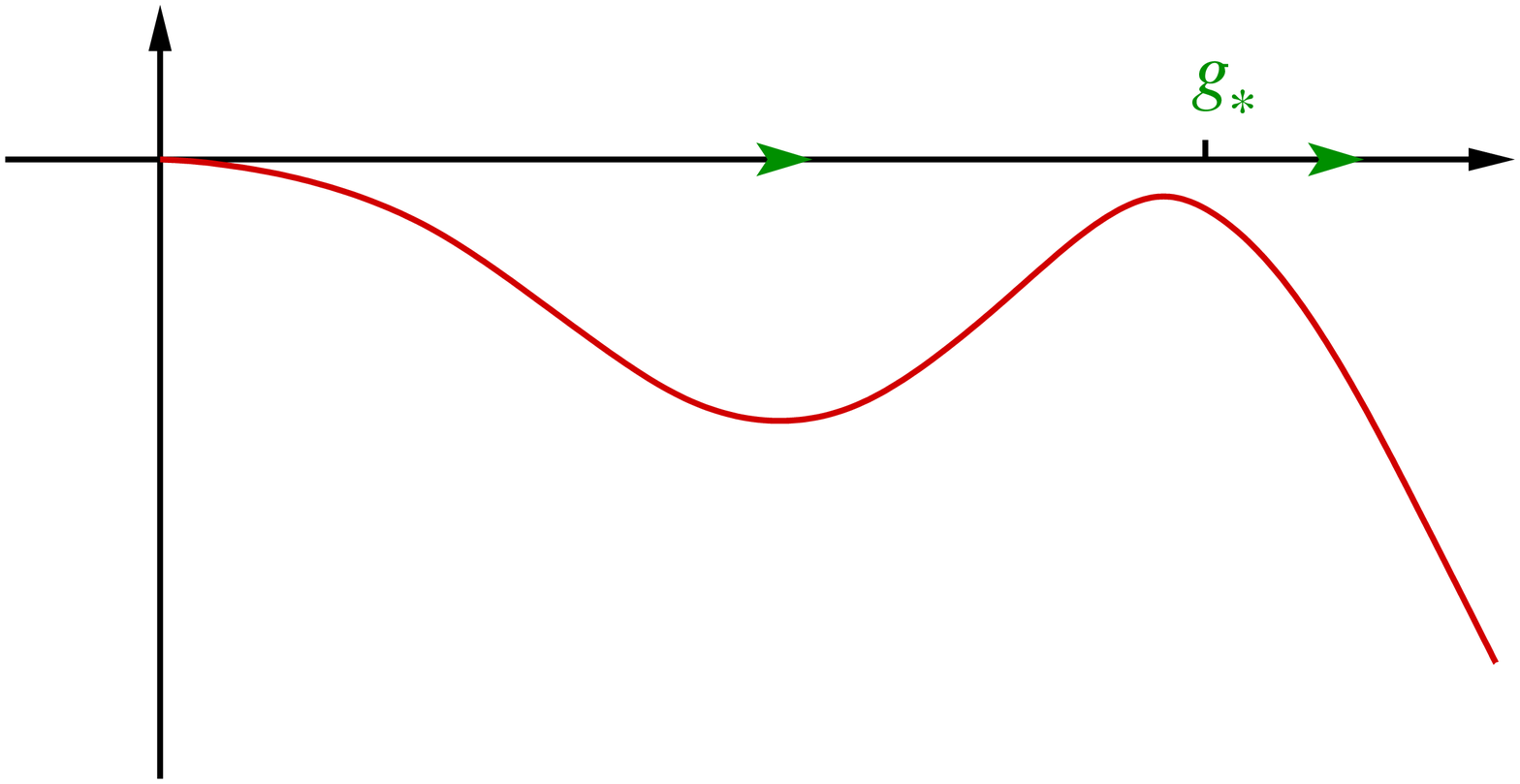}
\end{center}
\caption{Sketch of the beta function if the two-loop zero $g_*$ occurs at weak coupling (left) or at strong coupling (right).  The green arrows on the axis are the directions of the IR flow. The shape of $\beta(g)$ on the right indicates walking.
\label{fig1}}
\end{figure}

Let me begin by reviewing the possibilities raised by the two-loop beta function, as described by Banks and Zaks \cite{BZ}.
The perturbative expansion is
\begin{equation}
\beta(g^2)=-\frac{b_1}{16\pi^2}g^4-\frac{b_2}{(16\pi^2)^2}g^6+\cdots,
\end{equation}
where $b_1>0$ and $b_2<0$.
If we truncate at these two terms, the formula give an IR-attractive fixed point (IRFP) at $g=g_*$, as shown in Fig.~\ref{fig1}.

If $g_*$ is a weak coupling, then the two-loop calculation may be reliable. Flow into the IRFP implies that the massless theory possesses conformal dynamics at large distances.
This means that there is no confinement, no chiral condensate, and indeed no particles whatsoever!%
\footnote{We refer here to Casher's argument \cite{Casher} that massless quarks cannot form bound states in a vector-coupled gauge theory unless chiral symmetry is spontaneously broken. All color-singlet Green functions will possess only cuts starting at $q^2=0$.}

If $g_*$ is a strong coupling, on the other hand, the chiral condensate will form before the IRFP is reached, so the quarks will become massive and decouple from the IR dynamics.  The beta function past that point, returning to that of the pure gauge theory, will remain negative all the way to $g=\infty$ and there will be no actual zero.
The sketch for this case in Fig.~\ref{fig1} shows the marginal possibility that the beta function hovers near zero as this decoupling takes place, which is what underlies the hypothesis of walking.

The two-loop beta function makes a {\em prima facie\/} case for the existence of an IRFP and thus places the theory in what is called the {\em conformal window\/} \cite{DS}.
As reviewed in \cite{DS}, however, a calculation with the Bethe-Salpeter equation (the ``ladder'' approximation) points to $\chi$SB; a conjecture based on supersymmetry \cite{RS}, on the other hand, puts the theory back in the conformal window.
A lattice determination of the beta function can resolve the matter.

\section{The Schr\"odinger Functional method}

The Schr\"odinger functional (SF) \cite{SF,SFlatt} is a well-known method for calculating the beta function of the theory via imposing a background field.
Following the method of SF calculations for QCD, we employ Wilson fermions rather than staggered so that boundary values $U^i_{xt}$ (the background field) can be set on single time slices at $t=0$ and $t=L$ on an $L^4$ lattice; Wilson fermions also give us better control over the number of fermion flavors.
We add a clover term to remove $O(a)$ discretization errors, and we fix the clover coefficient self-consistently via tadpole improvement.

Wilson fermions, of course, break chiral symmetry explicitly and one must fix $\kappa=\kappa_c$ to have a massless theory in the continuum limit.
We define the quark mass by the axial Ward identity,
\begin{equation}
m_q \equiv \frac12 \;
\frac{\partial_4 \eval{A_4^b(t)\; \co^b(t'=0,\vec{p}=0)}}
{\eval{P^b(t)\; \co^b(t'=0,\vec{p}=0)}} \;
\bigg|_{t=L/2},
\end{equation}
where $\co$ is an operator on the boundary at $t=0$ while $A_4$ and $P$ are the axial and pseudoscalar densities, measured at zero spatial momentum at the center of the lattice.
Tuning to $m_q=0$ fixes $\kappa=\kappa_c$.

As is usual in SF calculations, we give the fermion fields a spatial twist.  The boundary conditions then serve as an efficient IR cutoff, even stabilizing the fermion inversions at $\kappa=\kappa_c$ and allowing us to study the massless theory directly.

In the continuum limit, the background field depends only on the size $L$ of the system, so the method gives the running coupling at the IR scale, $g^2(L)$.
More precisely, one is to calculate the potential $\Gamma\equiv-\log Z$ and compare it to the classical Yang-Mills action $S^{cl}$ of the background field configuration, giving $g^2(L)$ via
\begin{equation}
\Gamma=\frac1{g^2(L)}S^{cl}.\label{action}
\end{equation}
In a lattice Monte Carlo calculation, however, one cannot calculate $\Gamma$ directly.
The trick is to let the boundary values $U^i$ depend on a parameter $\eta$. Then by differentiating Eq.~(\ref{action}) we relate $g^2(L)$ to the expectation value of an operator, viz.,
\begin{equation}
\frac{\partial \Gamma}{\partial{\eta}}=
\eval{\frac{\partial S_{gauge}}{\partial{\eta}}
-\tr \left( \frac1{D_F^\dagger}\;
\frac{\partial (D_F^\dagger D_F)}{\partial{\eta}}\;
\frac1{D_F} \right)}
= \frac{K}{g^2(L)}\,,
\qquad
K \equiv \frac{\partial S^{cl}}{\partial{\eta}}=37.7\ldots
\label{GF}
\end{equation}

To summarize:  In order to extract the scaling of the running coupling, we
\begin{enumerate}
\item fix the lattice size $L$ and the couplings $\beta$ and $\kappa=\kappa_c(\beta)$;
\item calculate $K/g^2(L)$ via the expectation value (\ref{GF}), and also 
\item calculate $K/g^2(2L)$ on a lattice twice as large.  The two lattices have the same bare parameters $(\beta,\kappa)$ and hence the same UV cutoff $a$.  We thus
\item obtain the discrete beta function (DBF), defined as the difference
\begin{equation}
B(u,2)=\frac{K}{g^2(2L)}-\frac{K}{g^2(L)}\,,
\end{equation}
which is a function of $u\equiv K/g^2(L)$.
\end{enumerate}

\begin{figure}[hb]
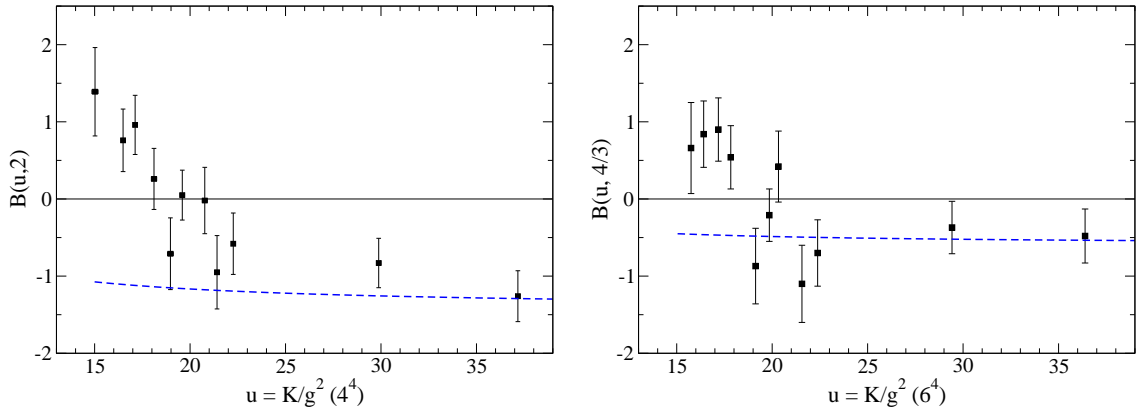

\begin{center}
\includegraphics[width=.48\textwidth,clip]{SSF7.eps}$\quad$
\includegraphics[width=.48\textwidth,clip]{SSF9.eps}
\end{center}
\caption{Left: Discrete beta function $B(u,2)$ obtained by comparing lattices of size $4^4$ and $8^4$.
Right: $B(u,4/3)$ obtained from lattices of size $6^4$ and $8^4$.
The dashed curves are the two-loop predictions.
\label{fig2}}
\end{figure}

We show the result of this procedure in Fig.~\ref{fig2} (left).
The DBF $B(u,2)$, obtained from lattices of size $4^4$ and $8^4$, evidently crosses zero at $g^2\simeq2.0$.  This is a much weaker coupling
than $g^2\simeq10$ as found in two-loop perturbation theory.  (For comparison, the two-loop DBF is plotted as a dashed curve.) On the face of it, this result demonstrates that the massless theory possesses an IRFP and hence that the IR theory is conformal.

If a lattice of size $4^4$ seems small, we can compare instead lattices of size $6^4$ and $8^4$, which yield the DBF $B(u,4/3)$ for the smaller scale factor $4/3$ (Fig.~\ref{fig2}, right).
As one might expect, the result is generally closer to zero than $B(u,2)$, and hence the error bars are relatively larger.  The crossing of zero is evident nonetheless.  We note that the data points in each plot are statistically independent, but the two plots are linked since the same $8^4$ data are used in each.

\section{\em Caveat cursor}

The DBF is, in principle, a continuum quantity that relates couplings at different IR scales $L$.
A lattice calculation of the DBF introduces an implicit dependence on the lattice spacing.
It goes without saying that our results contain lattice artifacts, which we cannot estimate since we have worked so far at a single lattice spacing for each value of $g$ and for each rescaling factor (2 and $4/3$).

Even when a satisfactory continuum limit is reached, however, one must ask whether the theory is really described by a single running coupling.
After all, in the IR regime any theory will need more than one term in its Lagrangian to describe its spectrum and interactions, and these terms will be generated by RG transformations.
Continuum perturbation theory automatically limits the effective Lagrangian to renormalizable couplings, and thus one always speaks of a single beta function for QCD or other (massless) gauge theories.
A lattice theory, on the other hand, can quickly generate many terms in the effective Lagrangian unless a physically reasonable truncation is used in the RG transformation.

\begin{figure}[htb]
\begin{center}
\includegraphics[width=.6\textwidth]{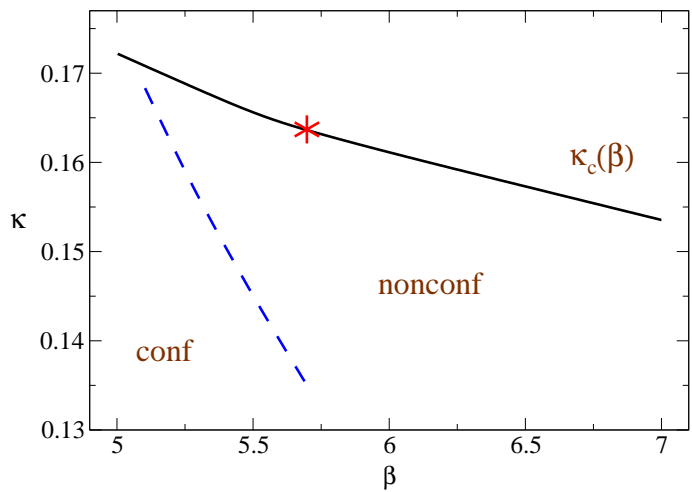}
\end{center}
\caption{Phase diagram in the $(\beta,\kappa)$ plane.  The solid curve is $\kappa_c(\beta)$, where the AWI quark mass $m_q$ vanishes.  The star marks the couplings corresponding to the IRFP of the SF effective coupling. 
The other curve denotes the finite-temperature/finite-volume phase transition for $L=8$. Presumably the finite-volume curve intersects the $\kappa_c$ curve, but we have not gone there.  See T.~DeGrand's talk~\cite{TDG} for more information on the phase diagram.
\label{fig3}}
\end{figure}

One necessary condition that the lattice theory be well described by a single coupling at scale $L$ is to verify that $L$ is too small for confinement to have set in.
Figure~\ref{fig3} is the phase diagram of the lattice theory.
It is clear that on our lattice the IRFP is found in the weak-coupling phase, meaning that the IR scale $L=8a$ is well within the confinement radius (if any).

\begin{figure}[htb]
\begin{center}
\includegraphics[width=.8\textwidth]{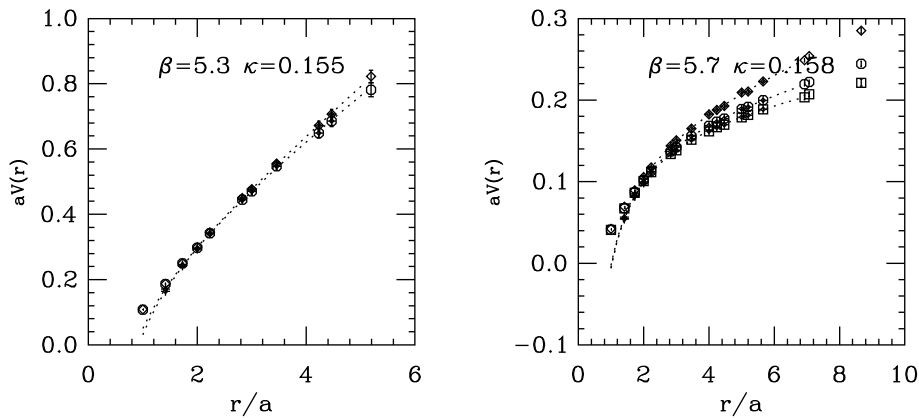}
\end{center}
\caption{Examples of the $q\bar q$ potential from our data. On the left, the string tension is
large; the Sommer parameter $r_0/a$ is
 measurable and we can obtain a good fit to $V(r)$. On the right, the string tension shrinks away as the temporal extent of the Wilson loop grows, and we cannot perform a reliable fit to $V(r)$.
The plot on the right represents the situation near the fixed point in Fig.~\protect\ref{fig3}. The lattice sizes are $8^3\times12$ (left) and $12^4$ (right).
\label{fig4}}
\end{figure}

If we examine this more closely, the simplicity of a gauge theory at a given scale may be judged by the behavior the $q\bar q$ potential.
If the potential is almost Coulomb, meaning $V(r)= g^2(r)/r$ with a coupling $g$ that varies only slowly with $r$, then one may proceed as if $g$ is the only coupling.  Again, we expect that this breaks down at large distances in QCD, where $V(r)$ first becomes linear in $r$ as the confining flux tube forms and then decays exponentially as it breaks.  It may be seen in Fig.~\ref{fig4} that when we measure $V(r)$ we find it to be consistent with a Coulomb potential in the neighborhood of the couplings corresponding to the IRFP.

Confirmation of the IRFP will thus come from (1) checking the DBF with more and larger volumes, (2) further understanding of the phase diagram in the $(\beta,\kappa)$ plane, and eventually (3) understanding of the continuum limit.
All this is in progress.
Then the challenge will be to measure properties of the conformal theory at the fixed point.  This will entail at least the calculation of operator exponents, which will govern how the fixed point is approached from the theories in its basin of attraction and from nearby massive theories.

More work on this model has been presented by Daniel Nogradi at this conference \cite{Nogradi}.
This work was supported in part  by the Israel Science
Foundation under grant no.~173/05 and by the US Department of Energy. Our computer code is based on version 7 of the publicly
available code of the MILC collaboration \cite{MILC}.


\begin{thebibliography}{99}
\bibitem{Nelson}
  A.~E.~Nelson,
  %``Lattice calculations for physics beyond the standard model,''
  PoS {\bf LAT2006}, 016 (2006).
  %%CITATION = POSCI,LAT2006,016;%%

\bibitem{SSD}
  Y.~Shamir, B.~Svetitsky and T.~DeGrand,
  %``Zero of the discrete beta function in SU(3) lattice gauge theory with color
  %sextet fermions,''
  Phys.\ Rev.\  D {\bf 78}, 031502 (2008)
  [arXiv:0803.1707 [hep-lat]].
  %%CITATION = PHRVA,D78,031502;%%

\bibitem{Caswell}
  W.~E.~Caswell,
  %``Asymptotic Behavior Of Nonabelian Gauge Theories To Two Loop Order,''
  Phys.\ Rev.\ Lett.\  {\bf 33}, 244 (1974).
  %%CITATION = PRLTA,33,244;%%
  
\bibitem{BZ}
  T.~Banks and A.~Zaks,
  %``On The Phase Structure Of Vector-Like Gauge Theories With Massless Fermions,''
  Nucl.\ Phys.\  B {\bf 196}, 189 (1982).
  %%CITATION = NUPHA,B196,189;%%

\bibitem{Belyaev}
  A.~Belyaev, R.~Foadi, M.~T.~Frandsen, M.~Jarvinen, A.~Pukhov and F.~Sannino,
  %``Technicolor Walks at the LHC,''
  arXiv:0809.0793 [hep-ph].
  %%CITATION = ARXIV:0809.0793;%%

\bibitem{Kogut}
  J.~B.~Kogut, J.~Polonyi, H.~W.~Wyld and D.~K.~Sinclair,
  %``Hierarchical mass scales in lattice gauge theories with dynamical light fermions,''
  Phys.\ Rev.\ Lett.\  {\bf 54}, 1980 (1985).
  %%CITATION = PRLTA,54,1980;%%

\bibitem{Casher}
  A.~Casher,
  %``Chiral Symmetry Breaking In Quark Confining Theories,''
  Phys.\ Lett.\  B {\bf 83}, 395 (1979).
  %%CITATION = PHLTA,B83,395;%%

\bibitem{DS}
  D.~D.~Dietrich and F.~Sannino,
  %``Walking in the SU(N),''
  Phys.\ Rev.\  D {\bf 75}, 085018 (2007)
  [arXiv:hep-ph/0611341].
  %%CITATION = PHRVA,D75,085018;%%

\bibitem{RS}
  T.~A.~Ryttov and F.~Sannino,
  %``Supersymmetry Inspired QCD Beta Function,''
  arXiv:0711.3745 [hep-th].
  %%CITATION = ARXIV:0711.3745;%%

\bibitem{SF}
  M.~L\"uscher, R.~Narayanan, P.~Weisz and U.~Wolff,
  %``The Schrodinger functional: A Renormalizable probe for nonAbelian gauge
  %theories,''
  Nucl.\ Phys.\  B {\bf 384}, 168 (1992)
  [arXiv:hep-lat/9207009];\\
  %%CITATION = NUPHA,B384,168;%%
  M.~L\"uscher, R.~Sommer, P.~Weisz and U.~Wolff,
  %``A Precise determination of the running coupling in the SU(3) Yang-Mills
  %theory,''
  {\em ibid.} {\bf 413}, 481 (1994)
  [arXiv:hep-lat/9309005];\\
  %%CITATION = NUPHA,B413,481;%%
  S.~Sint,
  %``On the Schrodinger functional in QCD,''
  {\em ibid.} {\bf 421}, 135 (1994)
  [arXiv:hep-lat/9312079];
  %%CITATION = NUPHA,B421,135;%%
  %``One Loop Renormalization Of The QCD Schrodinger Functional,''
  {\bf 451}, 416 (1995)
  [arXiv:hep-lat/9504005];\\
  %%CITATION = NUPHA,B451,416;%%
  M.~Della Morte {\em et al.} [ALPHA Collaboration],
  %``Computation of the strong coupling in QCD with two dynamical flavours,''
  {\em ibid.} {\bf 713}, 378 (2005)
  [arXiv:hep-lat/0411025], and references therein.
  %%CITATION = NUPHA,B713,378;%%
  
\bibitem{SFlatt}
  T.~Appelquist, G.~T.~Fleming and E.~T.~Neil,
  %``Lattice Study of the Conformal Window in QCD-like Theories,''
  Phys.\ Rev.\ Lett.\  {\bf 100}, 171607 (2008)
  [arXiv:0712.0609 [hep-ph]];\\
  %%CITATION = PRLTA,100,171607;%%
E.~T.~Neil, these proceedings.

\bibitem{TDG}
T. DeGrand, these proceedings:
  T.~DeGrand, Y.~Shamir and B.~Svetitsky,
  %``Exploring the phase diagram of sextet QCD,''
  arXiv:0809.2953 [hep-lat].
  %%CITATION = ARXIV:0809.2953;%%

\bibitem{Nogradi}
D.~Nogradi, these proceedings.

\bibitem{MILC} {\tt http://www.physics.utah.edu/\%7Edetar/milc/}

\end{thebibliography}
\end{document}